\documentclass[11pt]{article}

\usepackage{moriond,epsfig}

\def\NPB{{\em Nucl. Phys.}}
\def\PLB{{\em Phys. Lett.}}
\def\PRL{{\em Phys. Rev. Lett.}}
\def\PRD{{\em Phys. Rev.}}
\def\ZPC{{\em Z. Phys.}}
\def\EPJ{{\em Eur. Phys. J.}}

\newcommand{\beq}{\begin{equation}}
\newcommand{\eeq}{\end{equation}}
\newcommand{\beqn}{\begin{eqnarray}}
\newcommand{\eeqn}{\end{eqnarray}}
\newcommand{\beqns}{\begin{eqnarray*}}
\newcommand{\eeqns}{\end{eqnarray*}}
\newcommand{\bec}{\begin{center}}
\newcommand{\eec}{\end{center}}

\newcommand{\hm}{\hspace{-0.05cm}}

\newcommand{\intl}{\int\limits}
\newcommand{\ointl}{\oint\limits}
\newcommand{\mc}{\multicolumn}

\def\Cl{Collaboration}
\def\tauto{$\tau^{-\!}\rightarrow\,$}
\def\nut{$\,\nu_\tau$}

\def\ee{$e^+e^-$}

\def\daqedhZ{$\Delta\alpha_{\rm had}(M_{\rm Z}^2)$}
\def\amuhad{$a_\mu^{\rm had}$}

\def\ie{{\em i.e.}} 
\def\eg{{\em e.g.}} 
\def\ea{{\em et al.}} 
\def\via{via} 

\begin{document}
\pagestyle{empty}
\vspace*{-1.8cm}
\begin{flushright}{\bf LAL 01-74}\\
{October 2001}\\
\end{flushright}
\vskip 3.5 cm
\title{\Large HADRONIC CONTRIBUTION TO \boldmath$(g-2)_\mu$\vspace*{1cm}}

\author{\bf\Large Andreas H\"ocker \vspace*{0.5cm}}

\address{{\large\bf Laboratoire de l'Acc\'el\'erateur Lin\'eaire,}\\
	IN2P3-CNRS et Universit\'e de Paris-Sud, BP 34, 
	F-91898 Orsay Cedex, France\vspace*{1.5cm}}

\maketitle\abstracts{The recent precise measurement of the muon 
magnetic anomaly $(g-2)_\mu$ at BNL opens a window into 
possible new physics, provided the contribution from hadronic 
vacuum polarization is well understood. This talk summarizes the
development in the evaluation of the leading order hadronic 
contributions. Significant improvement has been 
achieved in a series of analyses which is presented 
historically in three  steps: (1), use of $\tau$ spectral 
functions in addition to \ee\ cross sections, (2), extended
use of perturbative QCD and (3), application of QCD sum rule 
techniques. The uncertainties, in particular concerning the
CVC hypothesis used in step (1), and global quark-hadron duality
employed in steps (2) and (3) are discussed. No new analysis results
are given in these proceedings.
\vspace*{6cm}}

\begin{center}
{\it Talk given at the ``XXXVIth Rencontres de Moriond'':\\ Electroweak Interactions and Unified Theories,\\
Les Arcs (France), March 10-17, 2001}
\end{center}
\newpage
\pagestyle{plain}

%
%
\section{Introduction}

Precision measurements of electroweak observables provide powerful tests 
of the Standard Model. In the last 10 years significant progress
has been achieved in this direction owing to the accurate and complete
results from the LEP, SLC and TEVATRON colliders. These measurements yielded
for the first time unique information from vacuum polarization effects
in weak boson propagators which allowed the mass of the Higgs boson to
be significantly bounded. At the other end of the energy scale, the
muon magnetic moment can now be measured with a precision such that new
physics can be probed, provided all the contributions from the Standard
Model can be under control.
In this talk, I shall discuss the new precise result on $(g-2)_\mu$ 
from the BNL experiment which deviates from the Standard Model
expectation: this exciting situation prompts us to critically examine 
the status of the theoretical prediction, in particular its most
delicate contribution from hadronic vacuum polarization. 
It turns out
that the same physics plays an important role in the analysis of
high-energy neutral-current data through the running of the 
electromagnetic coupling from $q^2=0$ to $M_Z^2$, relevant for 
limits on the Higgs mass.

%
%
\section{The muonic $(g-2)$}

\begin{figure}[h]
\bec
\hspace{0.0cm}\includegraphics[angle=0, width=0.14\textwidth]{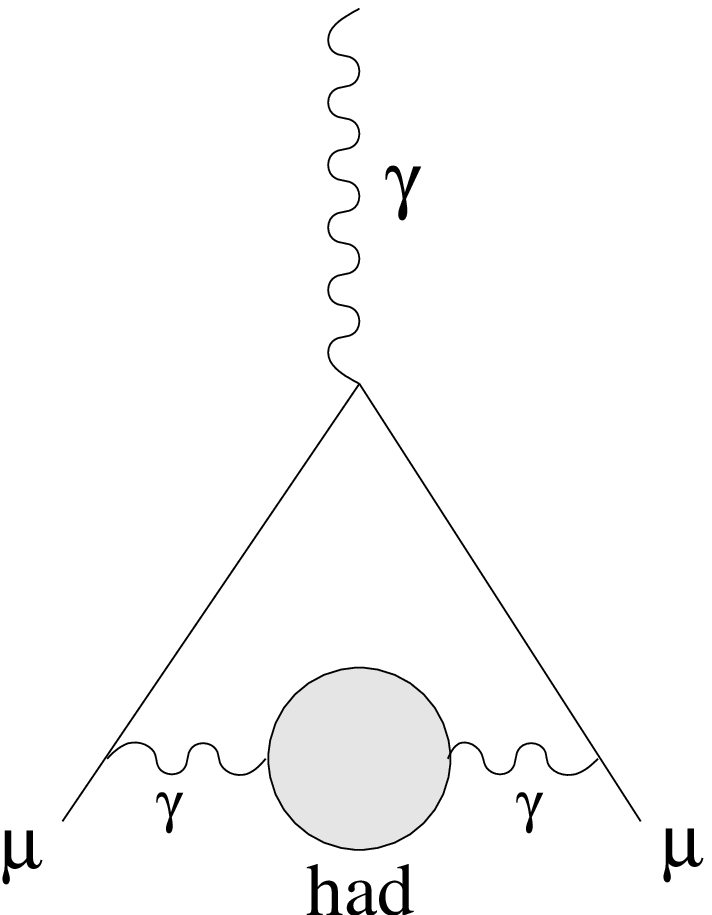}
\hspace{0.35cm}\includegraphics[angle=0, width=0.14\textwidth]{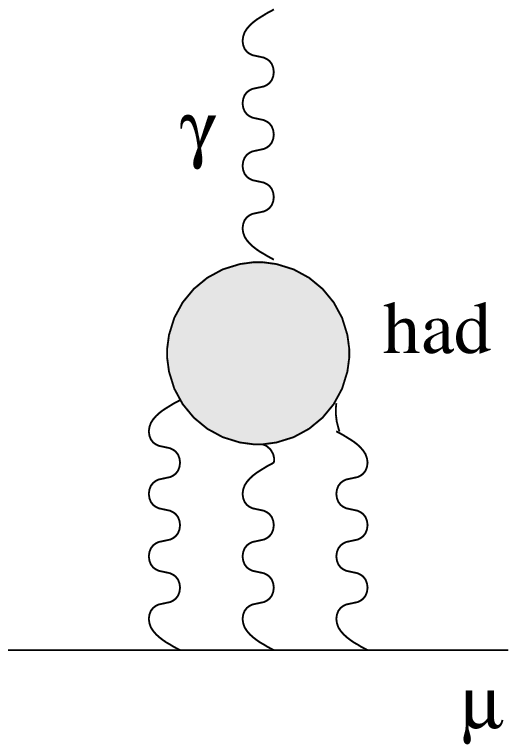}
\hspace{0.35cm}\includegraphics[angle=0, width=0.14\textwidth]{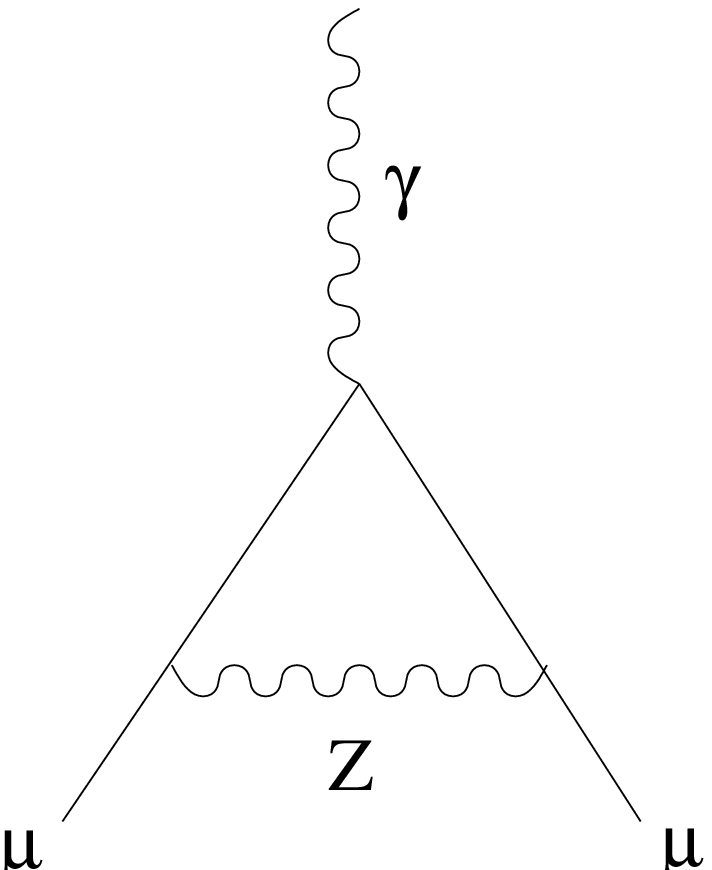}
\hspace{0.35cm}\includegraphics[angle=0, width=0.14\textwidth]{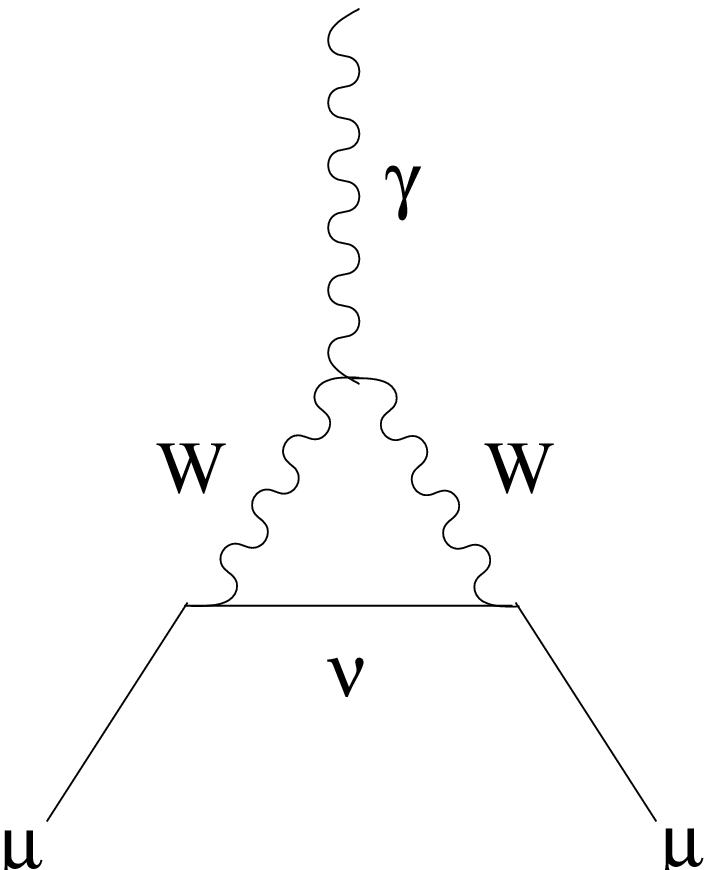}
\hspace{0.35cm}\includegraphics[angle=0, width=0.14\textwidth]{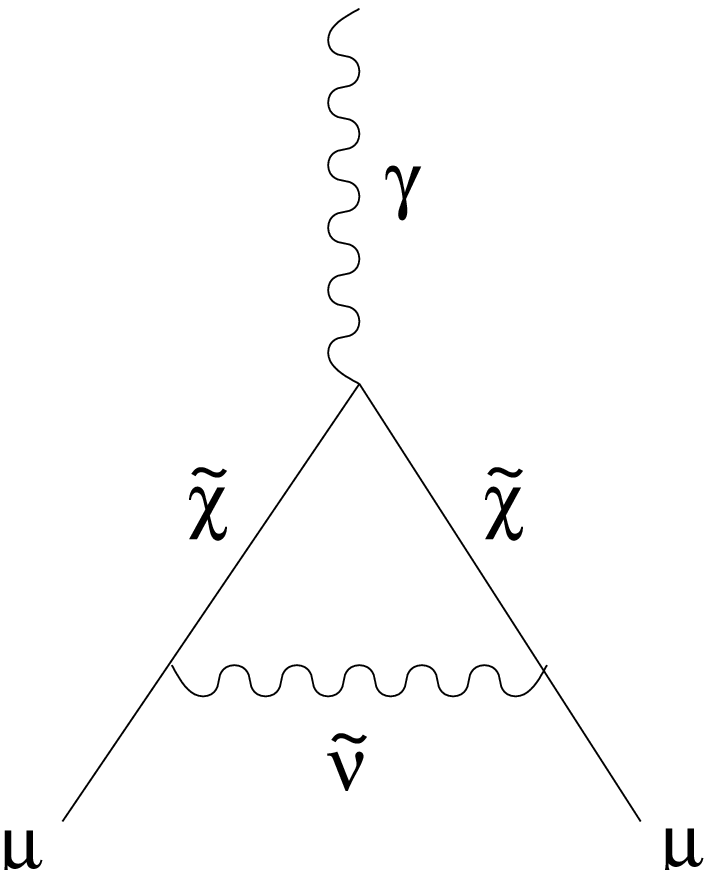} 
\hspace{0.35cm}\includegraphics[angle=0, width=0.14\textwidth]{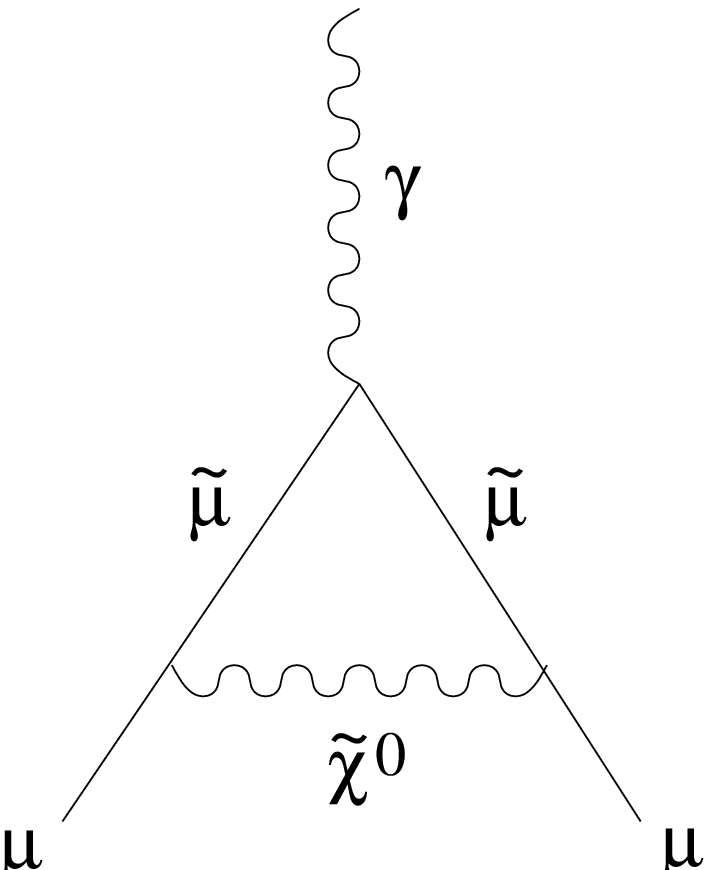}
\eec
\caption[.]{\label{fig:graphs}
Feynman diagrams corresponding to specific contributions
to $a_\mu$: first-order hadronic vacuum polarization, hadronic 
light-by-light scattering, first-order weak interaction and 
possible supersymmetric contributions.
}
\end{figure}

The muon magnetic anomaly $a_\mu$ receives 
contributions from all sectors of the Standard Model,
\beq
    a_\mu({\rm SM})\;\equiv\;\left(\frac{g-2}{2}\right)_\mu 
	\:=\: a_\mu^{\rm QED} + a_\mu^{\rm weak} + a_\mu^{\rm had}~,
\eeq
the dominant diagrams of which are depicted in Fig.~\ref{fig:graphs}. The pure QED contribution,
$a_\mu^{\rm QED}=116 584 705.7(2.9)\times10^{-11}$,
has been calculated to fourth order which represents
a {\em tour de force}, only performed by 
one group~\cite{kinoshita} (some slight change occurred recently due to 
computer precision problems in the original calculation~\cite{PCroberts}). 
The fifth order term has been 
estimated only, but was found to be small~\cite{qed5}. 
The weak contribution,
$a_\mu^{\rm weak} = 152(4)\times10^{-11}$,
is known to two-loops~\cite{weak2}. Large
logarithms of ${\rm ln}(M_W/m_f)$ occur, but can 
be resummed~\cite{peris}, leading to a robust prediction.
The contribution from hadrons stems mainly from vacuum polarization and
will be covered in the next section. Its absolute size 
$\simeq6800(160)$ ({\em ca.} 1995)
is such
that it must be known to a precision better than $1\%$ if the experiment
is to probe the level of the weak part. As it is well known, the 
first order correction from hadronic vacuum polarization 
(Fig.~\ref{fig:graphs}) cannot be calculated from ``first principles'' 
since most contributions arise from low-mass states, where 
quark confinement leads to resonances. Fortunately, 
the result can be expressed as a dispersion integral involving 
the total cross section for \ee\ annihilation into hadrons,
or alternatively its ratio $R(s)$ to the 
point-like cross section,
\beq
\label{eq:amuhad}
    a_\mu^{\rm had} 
           \,=\,\frac{\alpha^2}{3\pi^2}
           \intl_{4m_\pi^2}^\infty\!\!ds\,\frac{K(s)}{s}R(s)~,
\label{dispers}
\eeq
with $K(s)\sim m_\mu/s$, thus giving a 
large weight to the small $s$ region. An analog integral occurs 
for the running of 
$\alpha (Q^2)$, where $K(s)=(s-Q^2)^{-1}$. 
A small part of the hadronic contribution originates from the 
so-called light-by-light scattering (see Fig.~\ref{fig:graphs}). 
These diagrams
cannot be treated analogously and must be estimated through 
specific models for the hadron blob. As a consequence the result 
is less reliable and not known accurately, but not large compared 
to other contributions.
Obviously, the hadronic piece must be 
known more accurately, by a factor of at least three, before a precise
measurement can witness the effect of the weak interaction or a new
physics contribution of similar magnitude, such as Supersymmetry. This is
the motivation for an increased effort in the last few years to improve
the reliability and the accuracy of the hadronic contribution.

\begin{figure}[h]
\bec
\psfig{figure=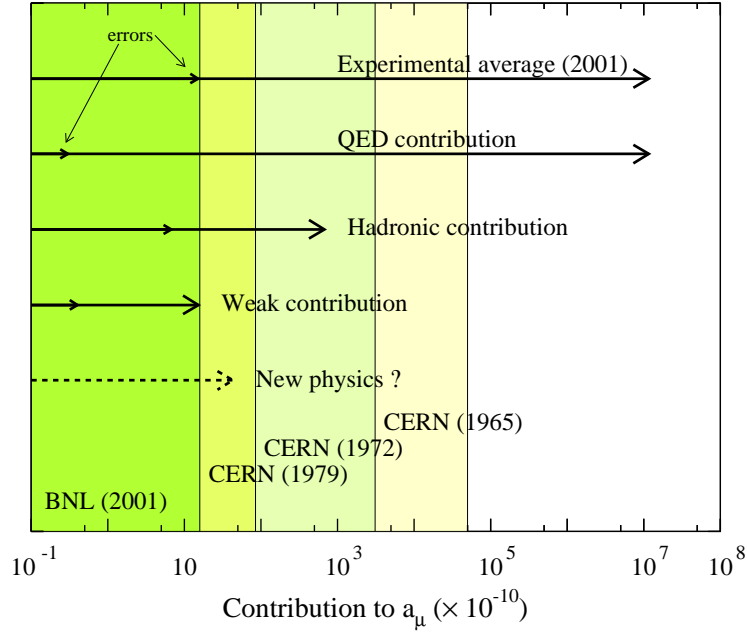,height=3.5in}
\eec
\caption{Experimental milestones on the precision of $a_\mu$ and 
	the levels of the different contributions and their 
	present	uncertainties (depicted by arrows) 
	expected in the Standard Model.}
\label{fig:expprog}
\end{figure}

The experimental progress on $a_\mu$ is chartered in Fig.~\ref{fig:expprog},
together with the levels of the different contributions 
expected in the Standard Model.
While the successive CERN experiments reach enough sensitivity to uncover 
the expected effect of hadrons, the program underway at BNL (E821) is now 
at the level of the weak contributions and reaches for
a four times smaller sensitivity, thus demanding a corresponding
improvement in the accuracy of the hadronic piece.  

\section{The precise BNL result}

The new value recently announced by E821~\cite{bnl01} has a precision 
three times higher than the previous combined CERN and BNL 
results~\cite{cern79,bnl00},
\beq
\label{eq:amubnl}
a_{\mu^+} = 11~659~202(16)~\times~10^{-10}~.
\eeq
The quoted uncertainty is dominated by statistics in muon decay counting
and the major systematic errors are estimated to $3.5\times10^{-10}$ 
for the precession frequency and $4.5\times10^{-10}$ for the magnetic 
field (NMR frequency).
The E821 experimenters compare their result to the expected SM value with
the hadronic contribution from vacuum polarization taken from Ref.~\cite{dh2}, 
\beq
a_{\mu^+}({\rm SM}) = 11~659~159.6(6.7)~\times~10^{-10}~.
\eeq
Averaging~(\ref{eq:amubnl}) with previous measurements yields
\beq
a_\mu({\rm exp})-a_\mu(\rm{SM}) = 43(16)~\times~10^{-10}~,
\eeq
where the error is dominated by the statistical experimental error
(theoretical systematic errors have been added in quadrature). 
A discrepancy at the level of $2.6 \sigma$ is observed and the
agreement with the SM can be questioned\footnote
{
  	The interpretation of the discrepancy in terms of standard
	deviations is approximately valid here, since the error
	is dominated by experimental uncertainties from the 
	$a_\mu$ measurement and from the hadronic contribution,
	where Gaussian Bayesian priors have been used to account for
	the systematic experimental errors.
	This is common practise, not to be 
	mixed up with a treatment of theoretical parameters which 
	are not statistically distributed quantities, but whose 
	uncertainties are not of dominance here.
}. 

\section{New physics in $(g-2)_\mu$?}

A $2.6 \sigma$ deviation is not sufficient to claim for a real discrepancy.
It is however large enough to speculate about its origin in order to 
rightfully question the different ingredients which could be responsible
for the effect. Four possibilities can be considered:
\begin{itemize}
\item {\em A statistical fluctuation?} This will be hopefully resolved 
	soon since the E821 Collaboration is presently analyzing their 
	2000 data with a factor four larger statistics.
\item {\em A systematic effect?} It is unlikely, since the estimated 
	systematic uncertainty is 2.5 smaller than the statistical 
	error and the (unaccounted) systematics would have to be 
	six times larger than the uncertainty of the systematics 
	accounted for.
\item {\em An error in the SM theory value?} The QED and electroweak 
	parts are under control at a level of two orders of magnitude 
	smaller than the observed effect. It is then proper to 
	question the hadronic contributions and we turn to this point 
	in detail in the next section.
\item {\em New physics?} Although it is in my opinion too early to
	speculate on this chance, it is clear that many 
	possibilities exist, witnessed by the paper flood
	since the publication of the BNL 
	result (see, \eg, \cite{marcianoNP,paperflood}): 
	Supersymmetry, muon
	substructure, anomalous electroweak couplings, leptoquarks, 
	lepton flavour violation, etc.
\end{itemize}

%
%
\section{Hadronic vacuum polarization for $(g-2)_\mu$ --- 
	improvements in three steps}

Since 1995 several improvements have been applied to the calculations of
hadronic vacuum polarization in order to cope with incomplete or unprecise
\ee\ data. Although QCD predictions were always used at higher energies
($>40~{\rm GeV}$), it became clear that reliable predictions could be made 
at much lower values. Let me identify the following three steps:
\begin{itemize}
\item[(1)] Addition of precise $\tau$ data using CVC (see, \eg,~\cite{adh,narison})
\item[(2)] QCD predictions at lower energies (see, \eg,~\cite{martin,dh1,kuhnstein,erler})
\item[(3)] Constraints from QCD sum rules (see, \eg,~\cite{schilcher,dh2,cvetic})
\end{itemize}

\subsubsection*{(1) Adding precise $\tau$ data under CVC}

The Conserved Vector Current (CVC) hypothesis expresses invariance under SU(2)
of the electroweak currents. For the problem at hand it relates the isovector
vector electromagnetic and the weak hadronic currents, as occurring in \ee\
annihilation and $\tau$ decays. From the point of view of strong interactions
this corresponds to a factorization of the hadronic physics: hadrons (quark
pairs) are created from the QCD vacuum and the probability to produce 
hadrons with well-defined quantum numbers at a given
mass is expressed through spectral functions. At low energy we expect 
spectral functions to be dominated by resonances, while QCD should provide
a good description at sufficiently high energies. The corresponding energy 
scale must be determined from experiment. 
The $I=1$ vector spectral function $v(s)$ for the two-pion
channel is related to the corresponding \ee\ cross section and $\tau$
branching ratio and invariant mass spectrum:
\beqn
v_{\pi^+\pi^-}(s)
	&=&
	\frac{s}{4\pi\alpha^2}\,
	 \sigma (e^+e^-\longrightarrow \pi^+\pi^-) ~,
	\\
v_{\pi^\pm\pi^0}(s)
	&\propto&
\frac{B_{\pi^\pm\pi^0}}{B_e}
\frac{1}{N_{\pi^\pm\pi^0}}\frac{dN_{\pi^\pm\pi^0}}{ds}
\frac{m_\tau^2}{\big(1-s/m_\tau^2\big)^{\!2}
          \big(1+2s/m_\tau^2\big)}~.
\eeqn
Hadronic $\tau$ decays represent a clean environment to study hadron 
dynamics which is in many ways complementary to \ee\ annihilation:
\begin{itemize}
\item $\tau$ data have excellent absolute normalization, because the 
	relevant branching ratios have been measured at LEP with 
	high statistics, large acceptance and very small non-$\tau$ 
	background~\cite{taubr}. On the other hand, the 
	shape of the spectral functions is subject to bin-to-bin 
	corrections from resolution effects and handling of fake 
	photons produced by hadron interactions in the electromagnetic 
	calorimeter. Therefore, the determination of the spectral function 
	requires an unfolding procedure which has the effect of strongly 
	correlating the errors of adjacent bins.
\item \ee\ data have just about the opposite behaviour: the point-to-point
	normalization is excellent, whereas the systematic uncertainties are 
	highly correlated among the measurements. The overall normalization 
	is a delicate issue, because of radiative corrections and 
	systematic errors from acceptance and luminosity.
\end{itemize}
The vector and axial-vector spectral functions have been measured at 
LEP by ALEPH~\cite{aleph_vsf,aleph_asf} and OPAL~\cite{opal_sf}. 
Detailed QCD studies have been performed by both collaborations.

\subsubsection*{\em SU(2) breaking}

If the $\tau$ data is to be used in the vacuum polarization calculations,
that is we identify $v_{\pi^\pm\pi^0}(s)$ with  $v_{\pi^+\pi^-}(s)$, it
is mandatory to consider in detail the amount of CVC 
violation~\cite{adh,czyz,neufeld}. Isospin breaking is expected mainly 
from electromagnetic effects and it has to be corrected for the 
calculation of the integral~(\ref{eq:amuhad}). The sources considered 
in the analysis and their quantitative effects on \amuhad\ are given 
in Table~\ref{tab:cvcbreak}. The dominant contribution comes from 
short distance electroweak radiative corrections 
to the effective four-fermion coupling 
$\tau^-\rightarrow (d\bar{u})^-$\nut.
It can be absorbed into an overall multiplicative electroweak 
correction $S_{\mathrm{EW}}=1.0194$~\cite{sirlin,braaten},
while remaining perturbative electroweak corrections are of order 
$\alpha^n(m_\tau)\,{\mathrm{ln}}^n(M_{\mathrm Z}/m_\tau)~0.3^n$ which 
is safe to ignore. The sub-leading non-logarithmic short distance 
correction, calculated to order $O(\alpha)$ at quark level~\cite{braaten},
$5\alpha(m_\tau)/12\pi\simeq0.0010$, is also small.
Additional intermediate-distance corrections have been computed only
for \tauto$\pi^-$\nut\ and the overall correction was 
found to be dominated by the leading 
logarithm from the short distance contribution~\cite{decker94}. 
The electromagnetic $\pi^\pm-\pi^0$ mass splitting affects the 
measured cross section through phase space corrections. Electromagnetic
corrections also affect the pion form factor, in particular the width 
of the $\rho$ resonance(s): the $\rho-\omega$ mixing, not present
in $\tau$ decays; the $\pi^\pm-\pi^0$ and 
$\rho^\pm-\rho^0$ mass splitting; electromagnetic decays.
The occurrence of second class currents is expected to be proportional
to the mass splitting-squared of the light $u,\,d$ quarks which
is negligible. We observe that most of the effects cancel, 
so that the net correction applied 
corresponds to approximately the pure short-distance
radiative correction $S_{\rm EW}$. It is important to further
investigate isospin-violating contributions to keep the
increasing precision of the $\tau$ data exploitable for the
purpose discussed here.

The use of $\tau$ data improves the precision on the evaluation 
of \amuhad\ by a factor of 1.6~\cite{adh}.

\begin{table}[h]
\setlength{\tabcolsep}{1.5pc}
\caption[.]{\label{tab:cvcbreak}Corrections for isospin violation applied to 
$\tau^-\rightarrow\pi^-\pi^0\nu_\tau$. }
\vspace{0.2cm}
\bec
{\small
\begin{tabular}{ll} \hline
 & \\[-0.35cm]
Source 	& $\Delta a_\mu^{\rm had}$ ($\times10^{-10}$) \\[0.1cm]\hline
 & \\[-0.35cm]
Radiative corrections to $\tau$ decays &  $-9.6\pm2.0$ \\[-0.05cm]
 {\footnotesize $S_{\rm EW} = 1.0194\pm0.0040$}   
 {\footnotesize\cite{sirlin,braaten}} & \\[-0.05cm]
 \mc{2}{l}{{\footnotesize (weak final state dependence expected,} 
 {\footnotesize verified for $\tau^-\rightarrow\pi^-\nu_\tau$)}
 {\footnotesize \cite{decker94}}}  \\[0.1cm]
Pion velocity: $\beta_- \ne \beta_0$  
& $-7.1$\\[-0.05cm]
 \mc{2}{l}{\footnotesize (due to EM $\pi^-$-$\pi^0$ mass splitting: 
affects cross section)}  \\[0.1cm]
 Form Factor \cite{adh,czyz}\small: & \\[0.0cm]
\hspace{0.5cm}{ $\circ$ $\rho$-$\omega$ interference}
	& $+3.7\pm0.6$ \\
\hspace{0.5cm}{ $\circ$ $\pi^-$-$\pi^0$ mass splitting  
		}{\footnotesize(affects $\Gamma_\rho$)}
	& $+3.3$ \\
\hspace{0.5cm}{ $\circ$ $\rho^-$-$\rho^0$ mass splitting  
	}{\footnotesize(affects $\Gamma_\rho$)}
	& $0\pm0.2$ \\
\hspace{0.5cm}{ $\circ$ EM $\rho$ decays: $\pi\gamma$, $\eta\gamma$,
	$\ell^+\ell^-$, $\pi\pi\gamma$ 
	}{\footnotesize(affect $\Gamma_\rho$)}
	&  $-0.2\pm1.2$\\
\hspace{0.5cm}{ $\circ$ second class currents}
	&  negligible \\[-0.05cm]
\hspace{1.0cm} {\footnotesize \eg,
$\tau^-\rightarrow\pi^-\eta\nu_\tau\propto (m_u-m_d)^2\sim10^{-5}$}
	& \\[0.1cm]\hline
 & \\[-0.35cm]
 Total correction 
	& $-9.9\pm2.4$ \\[0.1cm]\hline
\end{tabular}
}
\eec
\end{table}

\subsubsection*{(2) Replacing poor data by QCD prediction}

The data driven analysis~\cite{eidelman,adh} shows that 
to improve the precision on the dispersion integral,
a more accurate determination of the hadronic cross section between 
2~GeV and 10~GeV is needed, where some poorly measured and sparse 
data points dominate the final error. Indeed, QCD analyses 
using $\tau$ spectral functions~\cite{aleph_asf,opal_sf}
revealed the excellent applicability of the {\em Operator Product 
Expansion} (OPE)~\cite{wilson,svz} at the scale of the $\tau$ 
mass, $m_\tau\simeq1.8~{\rm GeV}$, and below. The OPE organizes perturbative 
and nonperturbative contributions to a physical observable through the
concept of global quark-hadron duality. Using moments of spectral 
functions, dimensional nonperturbative operators contributing to 
the $\tau$ hadronic width have been determined experimentally and 
found to be small. The evolution to lower energy scales proved (to 
some surprise) the validity of the OPE down to about 1.1~GeV.

An analog analysis based on spectral moments of \ee\ cross section 
measurements has been performed in Ref.~\cite{dh1} (and 
more recently in Ref.~\cite{menke} --- yielding compatible results).
The theoretical prediction of these moments and of the total 
hadronic cross section in \ee\ annihilation, $R(s_0)$, at a given 
energy-squared, $s_0$, involves the Adler $D$-function~\cite{adler}, 
related to the former \via\
\beq
\label{eq:radler}
     R(s_0) = \frac{1}{2\pi i}
              \hm\ointl_{|s|=s_0}\hm\hm\frac{ds}{s} D(s)~.
\eeq
Massless perturbative QCD predictions of $D$ are 
available~\cite{3loop} to order $(\alpha_s/\pi)^3$
(see also Ref.~\cite{evenmoreloops} for a heroic effort
to go beyond this).
The OPE of $D$ also includes second order quark 
mass corrections far from the production threshold~\cite{kuhn1} and
the first order dimension $D=4$ nonperturbative term
involving the gluon condensate, $\langle(\alpha_s/\pi) GG\rangle$, 
and the quark condensates, $\langle m_f\bar{q_f}q_f\rangle$, for 
the quark flavours $f$ (see, \eg, Ref.~\cite{bnp}). 
The complete dimension $D=6$ and $D=8$ operators are parameterized 
phenomenologically using the saturated vacuum 
expectation values $\langle{\cal O}_6\rangle$ and 
$\langle{\cal O}_8\rangle$, respectively. 
The nonperturbative operators (with the exception of the 
quark condensates, which are obtained from PCAC relations)
are determined experimentally by means of a combined fit 
of the theoretical moments to data. It results in 
a very small contribution from the OPE power terms to the lowest moment 
at the scale of $1.8~{\rm GeV}$ (repeated and confirmed at $2.1~{\rm GeV}$).
This is in agreement with the findings from the $\tau$ 
analyses. 

The calculation accounts 
for theoretical uncertainties, including the dependence on the 
choice of the renormalization scheme and scale, the uncertainty 
on the strong coupling, the missing term $(\alpha_s/\pi)^4$ and 
the ambiguity between contour-improved and fixed-order perturbation 
theory (see Refs.~\cite{dh1,aleph_asf}). Additional sources are the 
uncertainties on the running quark masses and on the 
nonperturbative contributions.
In spite of the implicit assumption of local duality for 
the theoretical prediction of $R$, the evaluation of the dispersion 
integral~(\ref{eq:amuhad}) turns the duality globally, \ie, 
remaining nonperturbative resonance oscillations 
are averaged over the integrated energy spectrum\footnote
{
	A systematic uncertainty is introduced through the cut
	at explicitly 1.8~GeV so that non-vanishing oscillations 
	may give rise to a bias after integration. The associated 
	(small) systematic error is estimated
	by means of fitting different oscillating curves 
	to the data around the cut region~\cite{dh1} .
}.

\begin{figure}[h]
\bec
\psfig{figure=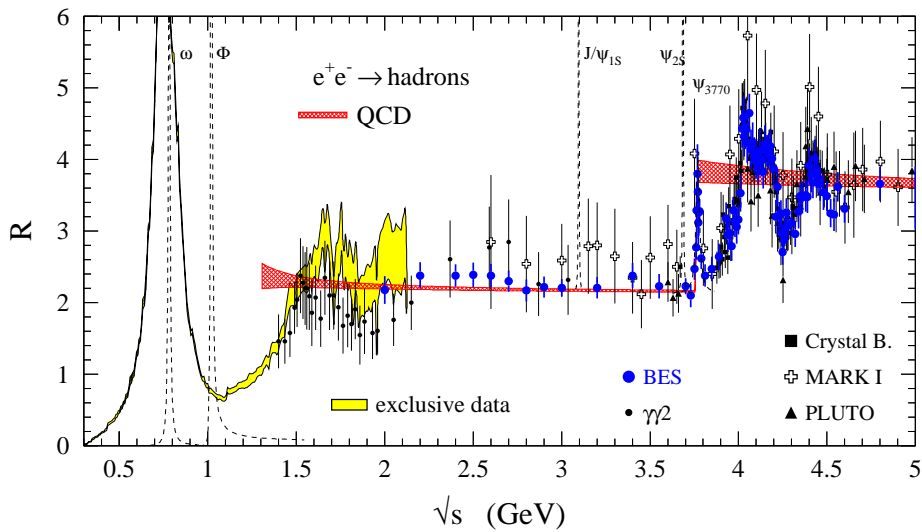,height=3.0in}
\eec
\caption[.]{The inclusive hadronic cross section ratio in \ee\
            annihilation versus the c.m. energy $\sqrt{s}$. 
            Shown by the cross-hatched band is the QCD prediction of 
	    the continuum contribution. The exclusive 
            \ee\ cross section measurements at low c.m. energies
            are taken from DM1, DM2, M2N, M3N, OLYA, CMD, ND and
            $\tau$ data from ALEPH (see~\cite{adh}
            for references and more detailed information).}
\label{fig:ree_bes}
\end{figure}

The available data points together with the theoretical prediction
(crossed hatched band) are shown in Fig.~\ref{fig:ree_bes}. Good
agreement is found between theory and the newest BES 
measurements~\cite{bes}, while older data are significantly 
higher. 

The preceeding discussion justifies the application of QCD 
predictions for $R$ between 1.8~GeV and the $D\bar D$ 
production threshold, as well as from 5~GeV up to infinity~\cite{dh1}. 
This yields a factor of 1.3 improvement on the precision
of \amuhad\ and a factor of 2.4 better accuracy on \daqedhZ. 
Similar precise analyses are performed in Ref.~\cite{kuhnstein} 
(applying a renormalization of experimental data on the $c\bar c$ 
resonances using QCD predictions of the continuum and assuming 
systematics to be correlated) and Ref.~\cite{erler} (see also 
Ref.~\cite{yndurain01} for a recent theory driven analysis).

\subsubsection*{(3) Improving data with QCD sum rules}

It was shown in Refs.~\cite{schilcher,dh2} that the previous 
determinations can be further improved by using finite-energy QCD 
sum rule techniques in order to access theoretically energy 
regions where perturbative QCD fails locally. In principle, the 
method uses no additional assumptions beyond those applied in 
the previous section. 
The idea is to reduce 
the data contribution to the dispersion integrals by subtracting 
analytical functions from the singular integration kernel in 
Eq.~(\ref{eq:amuhad}), and adding the subtracted part subsequently 
by using theory only. Two approaches have been applied in 
Ref.~\cite{dh2}: first, a method based on spectral moments is 
defined by the identity
\vspace*{0.5cm}
\beq
\label{eq:f}
      a_{\mu,\,[2m_\pi,\;\sqrt{s_0}]}^{\rm had}= 
          \intl_{4m_\pi^2}^{s_0}\!\!ds\,R(s)
	\left[\frac{\alpha^2K(s)}{3\pi^2s} - p_{n}(s)\right]
             + \frac{1}{2\pi i}\!\!\ointl_{|s|=s_0}\!\!\!\frac{ds}{s}
             \left[P_{n}(s_0) - P_{n}(s)\right] D_{uds}(s)~,
\eeq

\vspace*{0.5cm}
\noindent
with $P_{n}(s)=\int_0^sdt\,p_{n}(t)$.
The regular functions $p_{n}(s)$ approximate the kernel $K(s)/s$ in 
order to reduce the contribution of the non-analytic first integral 
in Eq.~(\ref{eq:f}), which
is evaluated using experimental data. The second integral in
Eq.~(\ref{eq:f}) can be calculated theoretically in the framework of 
the OPE. The functions $p_{n}(s)$ are chosen in order to reduce 
the {\em uncertainty} of the data integral which is not necessarily 
equivalent to a reduction of its contribution.
A second approach~\cite{dh2} involving local quark-hadron duality
uses the dispersion relation of the Adler $D$-function
\vspace*{0.5cm}
\beq
\label{eq_dispd}
    D_f(Q^2) = Q^2\!\!\intl_{4m_f^2}^\infty \!\!ds\,\frac{R_f(s)}{(s+Q^2)^2}~,
\eeq

\vspace*{0.5cm}
\noindent
for space-like $Q^2=-q^2$ and quark flavours $f$, to approximate
the integration kernel.
The theoretical errors of both approaches are evaluated 
in close analogy to the analysis presented in the previous Section.
The improvement in accuracy on the dispersion integrals obtained 
from these constraints is weak for \amuhad\ but valuable
for \daqedhZ.

\section{Results}

Table~\ref{tab:amures} shows the experimental and theoretical 
evaluations of \amuhad\ for the distinguished energy regions. 
Experimental errors between different lines are
assumed to be uncorrelated, whereas theoretical errors, but
those from the $c\bar{c}$ and $b\bar{b}$ thresholds which are quark mass
dominated, are added linearly. The combination of the theoretical
and experimental evaluations of the dispersion relation
yields the final result (see Table~\ref{tab:amures} for a note
concerning the theoretical error):
\vspace*{0.5cm}
\beqns
\label{eq_res3}
   a_\mu^{\rm had}[(\alpha/\pi)^2]
      &=& (692.4 \pm 5.6_{\rm exp} \pm 2.6_{\rm theo})\times10^{-10}~, \nonumber\\
   a_\mu^{\rm SM}
      &=& (11\,659\,159.6 \pm 5.6_{\rm exp} \pm 3.7_{\rm theo})\times10^{-10}~,
\eeqns

\vspace*{0.5cm}
\noindent
dominated by the contribution from the $\rho(770)$ resonance.
The total $a_\mu^{\rm SM}$ value contains the contributions
from non-leading order hadronic vacuum polarization~\cite{krause2,adh} 
$a_\mu^{\rm had}[(\alpha/\pi)^3]=(-10.0\pm0.6)\times10^{-10}$, and from
hadronic light-by-light scattering (LBLS)
for which the average of the results given in Refs.~\cite{kinolight,light2}
is used
$\langle a_\mu^{\rm had}[{\rm LBLS}]\rangle=(-8.5\pm2.5)\times10^{-10}$
(see Ref.~\cite{melnikov} for a recent critical review).

\begin{table*}[h]
\setlength{\tabcolsep}{1.5pc}
\caption[.]{\label{tab:amures}
            Contributions to \amuhad\ from the 
            different energy regions. The subscripts in the first column
            give the quark flavours involved in the calculation.}
\vspace{0.2cm}
\begin{center}
{\small
\begin{tabular}{lc} \hline \\[-0.33cm]
Energy~(GeV)
                              & $a_\mu^{\rm had}\times10^{10}$ 
\\[0.07cm]
\hline \\[-0.33cm]
$(2m_\pi$ -- $1.8)_{uds}$
                              & $634.3\pm5.6_{\rm exp}\pm2.1_{\rm theo}^{(*)}$ 
\\[0.07cm]
$(1.8$ -- $3.700)_{uds}$ 
                              & $33.87\pm0.46_{\rm theo}$     
\\[0.07cm]
$\psi(1S,2S,3770)_c$ 
$+~(3.7$ -- $5)_{udsc}$
                              & $14.31\pm0.50_{\rm exp}\pm0.21_{\rm theo}$
\\[0.07cm]
$(5$ -- $9.3)_{udsc}$    
                              & $6.87\pm0.11_{\rm theo}$ 
\\[0.07cm]
$(9.3$ -- $12)_{udscb}$
                              & $1.21\pm0.05_{\rm theo}$ 
\\[0.07cm]
$(12$ -- $\infty)_{udscb}$
                              & $1.80\pm0.01_{\rm theo}$ 
\\[0.07cm]
$(2m_t$ -- $\infty)_t$
                              & $\approx0 $ 
\\[0.07cm]
\hline\\[-0.33cm]
$(2m_\pi$ -- $\infty)_{udscbt}$
                              & $692.4\pm5.6_{\rm exp}\pm2.6_{\rm theo}$ 
\\[0.07cm]
\hline
\end{tabular}
}
{\footnotesize 
\parbox{14cm}
{
\vspace{0.2cm}
$^{*}\,$The theoretical error accounts for uncertainties 
	concerning the QCD prediction only. Due to the 
	correlated average procedure applied in Ref.~\cite{adh}, 
	uncertainties from CVC and radiative corrections are 
	folded into the systematic part of the experimental 
	error.
}}
\end{center}
\end{table*}

%
%
\section{Conclusions and perspectives}

\begin{figure}[h]
\bec
\psfig{figure=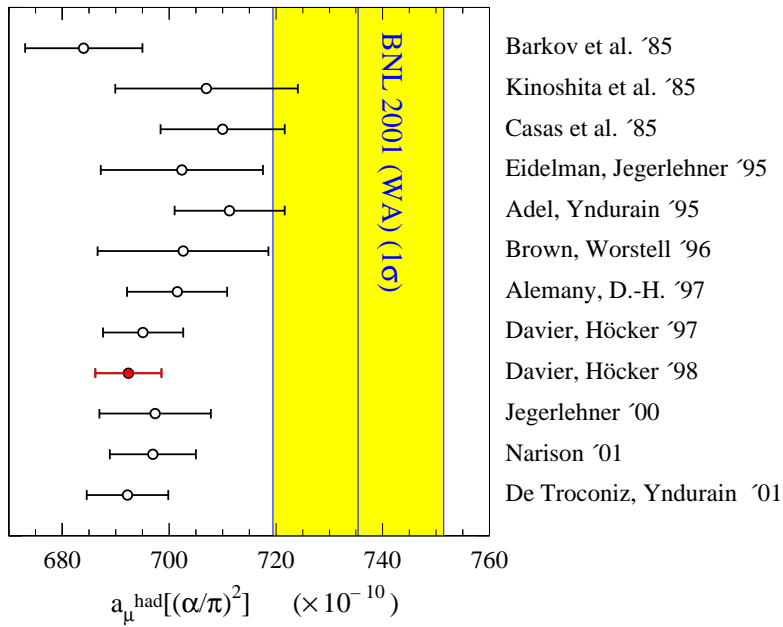,height=3.5in}
\eec
\vspace{-0.4cm}
\caption[.]{Comparison of lowest order $a_\mu^{\rm had}$ evaluations and the 
	experimental result corrected for the QED, weak and higher order
	hadronic contributions. 
	The theoretical values are taken from Refs.~\rm
            \cite{amuhadall,eidelman,adh,dh1,dh2,narison,yndurain01}.}
\label{fig:amu_res}
\end{figure}

Much effort has been undertaken during the last years to 
ameliorate the theoretical predictions on \amuhad. 
The currently most precise value obtained for the hadronic 
contribution is~\cite{dh2} 
$a_\mu^{\rm had}=(692.4 \pm 6.2)\times10^{-10}$. 
Figure~\ref{fig:amu_res} gives a chronological compilation of 
published results. 

\noindent
Fairly
good agreement is observed among the newest evaluations, so that 
the discrepancy between the BNL value and the Standard Model
varies between $2\sigma$ and $2.6\sigma$.
To maintain the sensitivity on interesting physics of the 
experimental improvements to be expected from BNL, more 
theoretical effort is needed. In particular, a better precision 
on \amuhad\ requires further studies of the following items.
\begin{itemize}
\item	Radiative corrections in \ee\ annihilation data
\item	SU(2) breaking: let me recall that the $\tau$ data not only 
	provide precise and in many ways complementary
	cross section measurements, but they also constitute
	a powerful cross check.
	The current \amuhad\ evaluation being wrong would require
	not only the \ee\ data to have unaccounted systematics, but 
	also that CVC violation is much larger than expected, since
	the \ee\ and $\tau$ data are mutually (fairly) compatible.
\item 	More experimental information. In particular, complementary
	\ee\ measurements from, \eg, new precision experiments,
	or analyses of radiative events using data from existing
	\ee\ factories. 
\end{itemize}
(Also the contribution from hadronic LBLS is of much importance but 
not the subject of this talk.)
It is obvious that using the $\tau$ data and CVC represents only an 
auxiliary remedy to cope with the lack of precision in the \ee\ 
measurements. Certainly, the preferable scenario would be to 
improve the latter well below the $1\%$ accuracy so that one does
not need to include the former.
Moreover in such a situation precise tests of CVC could be 
performed, providing insight into the interesting physics of
possible violations.

%
%
\section*{Acknowledgements}
I gratefully acknowledge the very fruitful and pleasant collaboration with
Michel Davier.

%
%

\end{document}